# Hydrogen-induced fast fracture in a 1.5 GPa dual-phase steel

*Rama Srinivas Varanasi[1,2], #Motomichi Koyama[1], Shuya Chiba[1,3], Saya Ajito[1], Eiji Akiyama[1]

[1] Institute for Materials Research, Tohoku University, 2-1-1 Katahira, Aoba-ku, Sendai, 980-8577, Japan

[2] Department of Materials Science & Metallurgical Engineering, Indian Institute of Technology, Hyderabad-502285, India

[3] Graduate School of Engineering, Tohoku University, 6-6-01-2 Aramaki Aza Aoba, Aoba-ku, Sendai, Miyagi, 980-8579 Japan

**Corresponding authors:**

*Rama Srinivas Varanasi: varanasiramasrinivas@gmail.com

#Motomichi Koyama: motomichi.koyama.c5@tohoku.ac.jp

**Abstract**

This study clarifies the hydrogen embrittlement (HE) behavior in a 1.5 GPa ferrite-martensite dual-phase (DP) steel. Hydrogen pre-charging (3.8 mass ppm diffusible hydrogen), followed by slow strain tensile testing ($10^{-4}$ $s^{-1}$), resulted in a brittle fracture at 900 MPa within the elastic regime. Fractographic studies indicated that surface crack initiation consists of intergranular and quasi-cleavage morphology; site-specific transmission electron microscopy (TEM) investigations revealed sub-surface secondary crack blunting by ferrite. A mixed-mode morphology consisting of ductile and brittle features was observed adjacent to crack initiation. It differs from the previous investigation of uncharged DP steel, wherein a predominant brittle fracture was observed. Following significant crack growth, the pre-charged specimen exhibited predominant brittle fracture; site-specific TEM and transmission Kikuchi diffraction studies revealed {100} ferrite cleavage cracking. Electron backscatter diffraction studies were performed on the cross-sectional cracks. We explain the HE via hydrogen-induced fast fracture mechanism. During loading, hydrogen diffuses to the prior austenite grain boundary, resulting in hydrogen-induced decohesion. Subsequent hydrogen diffusion to the crack tip promotes brittle fracture at high crack velocity (>$V_{crit}$). The high crack velocity effectively inhibits crack blunting via dislocation emission, ensuring sustained brittle crack growth even after hydrogen depletion at the crack tip, resulting in {100} ferrite cleavage cracking. Based on TEM observations, we explain the formation of river pattern features on the {100} cleavage surface.

**Graphical Abstract**

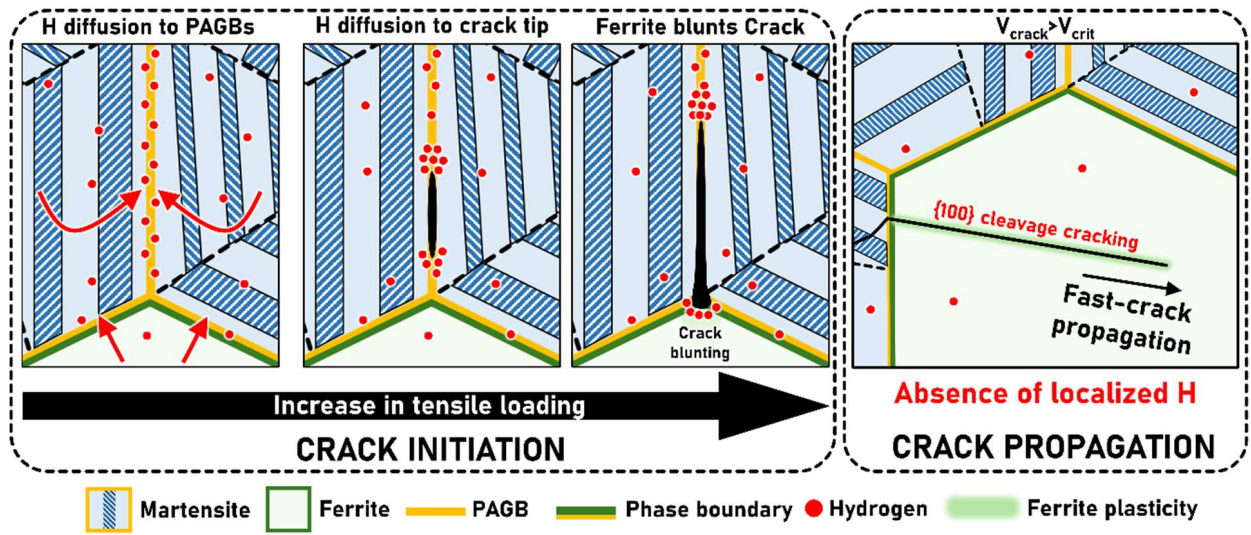

## 1  Introduction

Advanced high-strength steel (AHSS) consumption was estimated at 14 million tonnes in 2019 [1], and AHSSs accounted for up to ~2% of the global steel market in 2021 [2]. AHSSs are predominantly utilized in the transportation sector (which accounts for ~25 % of the global $CO_2$ emissions) [3]. The life cycle assessment of vehicles indicates that materials production and the operation phase are responsible for 8-32% and 63-92 % of the energy consumption, respectively [2]. Lightweighting can mitigate emissions associated with both production (by reducing the amount of AHSS required) and vehicle operation (due to decrease in the vehicle mass); the life-cycle energy savings can be up to 25 % [2].

While lightweighting using steels with a higher strength-to-weight ratio is an effective strategy, ensuring safety is paramount. The susceptibility to hydrogen embrittlement increases with an increase in strength [4–6]. Hydrogen embrittlement is a phenomenon wherein hydrogen in quantities of 0.04 wppm can render the AHSS brittle [7], exposing the structural steel to a risk of catastrophic failure [8,9]. To this end, the following alloy and microstructure design strategies have been devised to mitigate hydrogen embrittlement in AHSS:

1. Trapping of diffusible hydrogen via second-phase particles such as carbides, carbonitrides, and copper precipitates [10–16].
2. The propensity of hydrogen-induced grain boundary cracking is dependent on the structure of the boundary (or the parent austenite boundary in the case of martensitic steels) [17,18]. Hence, a grain boundary engineering (GBE) approach via the increase in the fraction of low-energy boundaries (or special boundaries) enhances resistance to hydrogen embrittlement [19,20].
3. Hydrogen-enhanced decohesion (HEDE) is one of the dominant hydrogen embrittlement mechanisms [21,22]. For example, in martensitic steels, hydrogen-induced intergranular cracking occurs along the prior austenite grain boundaries (PAGBs) [18,23,24]. Grain boundary segregation engineering (GBSE) via segregation of alloying elements such as Mo [25] and B [26] to the PAGBs reduces the susceptibility to hydrogen embrittlement.
4. Grain refinement, for instance, in martensitic steels [27–29], has been reported to mitigate hydrogen embrittlement.

Recently, Pinson et al. [30] designed martensitic steels with ferritic microfilm at the PAGBs to exploit the crack-arresting ability of ductile ferrite and achieved a 100% increase in the ductility in the absence of hydrogen. However, in the presence of hydrogen, at a slower deformation rate, it was observed that the ferritic film lost the ability to influence the crack propagation path, resulting in embrittlement [31]. However, the strategy of introducing ductile ferrite to mitigate hydrogen embrittlement is promising and merits further study, particularly in the context of ferrite-martensite dual-phase (DP) steels. This is because DP steels have a lean elemental composition, making them cost-effective, and can be industrially produced with a simple thermomechanical treatment [32].

Previously, we clarified the fracture mechanisms during uniaxial tensile loading in a 1.5 GPa DP steel consisting of 25% ferrite in the absence of hydrogen [33]. Despite 5% ductility, the DP steel exhibited predominant brittle fracture with ferrite undergoing {100} cleavage cracking [33]. It was attributed to the high local strain rate in ferrite caused by stress partitioning when a crack propagates from martensite to ferrite [33]. We performed U-bend tests to study the hydrogen-induced delayed fracture susceptibility [5], wherein it was observed that ferrite arrested both the surface and sub-surface crack propagation. Hence, it is important that we verify if ferrite can mitigate hydrogen embrittlement in a hydrogen-pre-charged specimen, which is the aim of the present work. We compare and contrast the micro mechanisms of delayed fracture in U-bend testing with that of H-assisted failure in pre-charged tensile specimen. We present a comprehensive understanding of the role of ferrite in the hydrogen embrittlement of the 1.5 GPa DP steel.

## 2 Methods and materials:

Table 1 shows the composition of the DP steel (in mass %) used in the present work. The detailed heat treatment process has been reported in our previous study [33]. The tensile specimens are 12.5 mm wide, have a gauge length of 60 mm, and are 1.6 mm thick; they correspond to the JIS 13B standard. Tensile tests were carried out in a Shimadzu Autograph AGS-X machine at strain rates of $10^{-3}$ s$^{-1}$ and $10^{-4}$ s$^{-1}$ for the hydrogen-charged specimen and at a strain rate of $10^{-3}$ s$^{-1}$ for the uncharged specimen. Electrochemical hydrogen charging was performed at a current density of 1 A m$^{-2}$ in an aqueous solution containing 3 wt.% NaCl and 3 g L$^{-1}$ NH$_4$SCN for 12 hours using a platinum counter electrode. The diffusible hydrogen content after electrochemical hydrogen charging was measured using a thermal desorption spectroscopy system (R-DEC Co., Ltd.) equipped with a quadrupole mass spectrometer. The heating rate was 100°C per hour.

Table 1 DP steel composition (in mass %)

| C | Si | Mn | P | S | Cr | Mo | Ni | Fe |
|---|---|---|---|---|---|---|---|---|
| 0.35 | 0.29 | 0.69 | 0.008 | 0.004 | 0.97 | 0.18 | 0.01 | Bal. |

Secondary electron (SE) imaging, electron channeling contrast imaging (ECCI), and electron backscatter diffraction (EBSD) were performed in a Carl Zeiss Merlin field emission scanning electron microscope (SEM) equipped with an EDAX Digiview 5 EBSD detector. EBSD investigations were conducted using a 20 kV and 10 nA beam. Specimens for EBSD and ECCI studies were prepared via mechanical polishing. After grinding the samples using SiC paper, they were polished using 3 µm monocrystalline diamond suspension from Buehler. It was followed by a 60 nm colloidal silica suspension. The EBSD data was analyzed using the OIM software, and points with a confidence index (CI) > 0.1 were considered.

Site-specific sample preparation for transmission electron microscopy (TEM) and transmission Kikuchi diffraction (TKD) studies were carried out in a dual-beam FEI Helios NanoLab 660i SEM-focused ion beam (FIB). Initially, ~200 nm of platinum was deposited via electron beam-assisted deposition to ensure no damage induced by the ion beam at the area of interest [34]. It was followed by ~2 µm of ion beam deposited platinum over it. To avoid ion-beam damage to TEM/TKD lamella during the TEM sample preparation, we incrementally reduced the voltage down to low-kV milling in the final step [35]. TKD measurements were performed in Carl Zeiss Merlin SEM with a step size of 20 nm using a TSL holder designed to be compatible with a standard EBSD detector. TEM investigations were conducted in JEOL ARM200F and JEOL JEM-2100F.

## 3 Results:

### 3.1 Initial microstructure

Fig. 1 is a representative microstructure of the DP steel, an overlay of the phase map with the corresponding image quality (IQ) map. As reported in the previous study, the ferrite area fraction is estimated to be ~25 % (based on IQ map of 80 µm X 120 µm) [33]. Additionally, we have the presence of MnS inclusions, as highlighted by the white circle in Fig. 1.

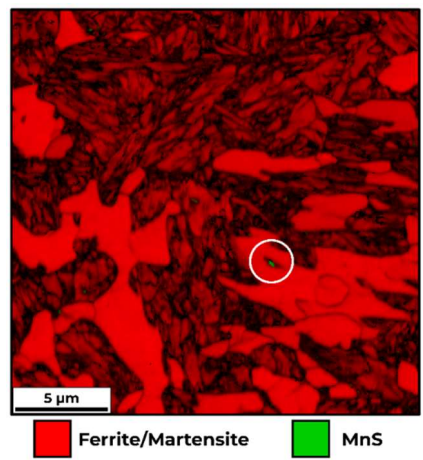

*Figure 1 Overlay of the phase map with image quality (IQ) map. The white circle highlights the MnS inclusion.*

### 3.2 Hydrogen embrittlement susceptibility

Fig. 2a shows the tensile properties of the DP steel in the uncharged condition and after hydrogen charging. The uncharged steel exhibited ultimate tensile strength (UTS) of 1.51 GPa and ductility (strain at failure) of 5.21% at a strain rate of $10^{-3}$ $s^{-1}$. In the presence of H, the sample failed at a UTS of 1.23 GPa and ductility of 1.13% when the tensile test was performed at a strain rate of $10^{-3}$ $s^{-1}$. At a strain rate of $10^{-4}$ $s^{-1}$, the hydrogen-charged specimen exhibited drastic hydrogen embrittlement, as fracture occurred in the elastic regime at a UTS of 0.90 GPa and ductility of 0.57%. Strain rate sensitivity is further discussed in section 4.4. Fig. 2b indicates the hydrogen desorption spectra. In the current work, the diffusible hydrogen content is defined as the total desorbed hydrogen from room temperature until 250°C where the first peak was present and was measured to be 3.8 mass ppm.

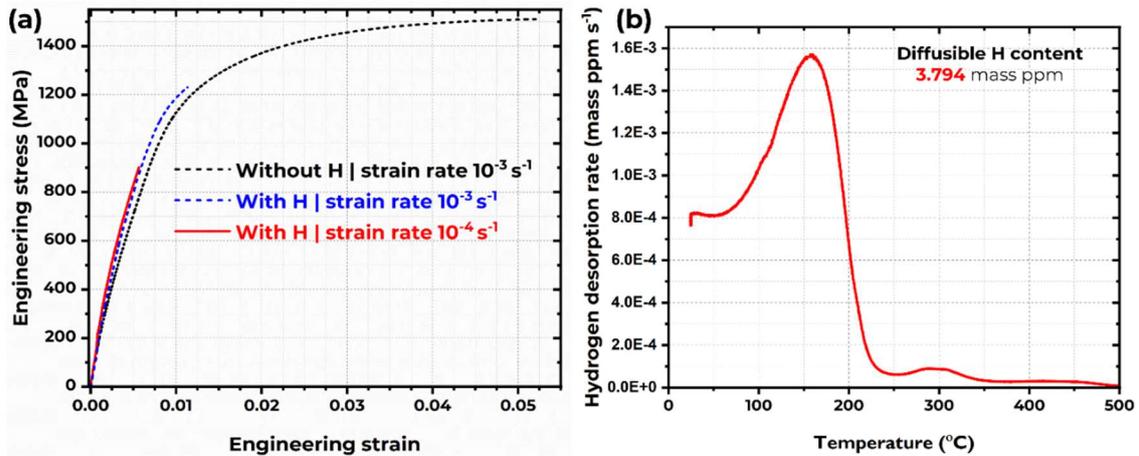

**Figure 2** *(a) Tensile properties of the uncharged and hydrogen-charged specimen. (b) Hydrogen desorption spectra of the hydrogen-charged specimen at a heating rate of 100°C per hour.*

### 3.3    Fractography

To understand the hydrogen-induced fracture, we performed detailed fractographic investigations. The fracture surface of the hydrogen-charged specimen, tensile-tested at a strain rate of $10^{-4}$ $s^{-1}$, is shown in Fig. 3a. We observed that a crack initiates adjacent to the surface of the specimen, gradually propagating to the center of the specimen (along the thickness direction of the sample), as indicated by the red arrow in Fig. 3a. Subsequently, the crack propagates from one side of the tensile specimen to the other (left to right in Fig 3a, as highlighted by red arrow). The faint chevron fracture markings in the fractograph further indicate that as the crack propagated from one end of the tensile specimen to another, crack propagation occurred from the center of the specimen to the surface (as shown by the two smaller red arrows in Fig. 3a). This is further confirmed by the presence of shear lip at the surface of specimen except at the region of crack initiation. Figs. 3b-c show fractographs at a higher magnification. Fig. 3c indicates the shear lip that corresponds to the red rectangular box in Fig. 3b. Detailed fractographs of the crack initiation region are provided in Fig. 4. Fractography of the area ahead of the crack initiation region, as the crack propagated from the surface towards the center (along the thickness), is shown in Fig. 5. Fig. 6 indicates the fracture surface upon significant propagation of the crack.

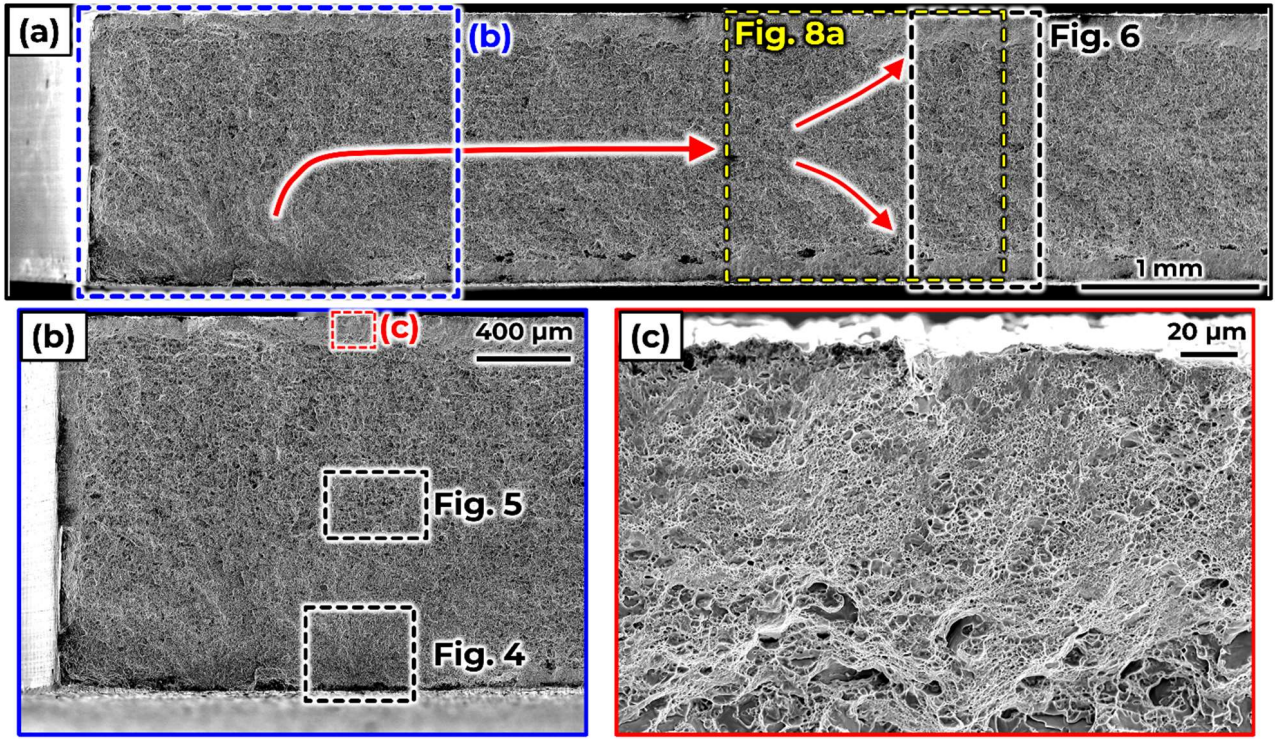

*Figure 3 (a) Fracture surface of the DP steel pre-charged with hydrogen after tensile testing at a strain rate of $10^{-4}$ $s^{-1}$. The red arrow indicates the crack propagation from the surface of the specimen towards the center (along the thickness). Subsequently, the crack propagates from one end of the tensile specimen to the other. The two smaller red arrows indicate crack propagation from the center of the thickness toward the surface of the specimen. (b) Fractograph at a higher magnification corresponding to the blue rectangular box in (a). (c) Micrograph indicating shear lip at the surface of specimen for a region that corresponds to the red rectangular box in (b). The black rectangular boxes correspond to Figs. 4-6, which display the fracture surface at a higher magnification.*

We observe features in the fracture surface in Fig. 4a propagating radially outwards from the crack initiation region, suggesting that the crack growth occurred radially immediately upon crack initiation. Higher magnification micrographs of the crack initiation region are shown in Figs. 4b-c. Intergranular fracture features are observed in Fig. 4c (red arrows). Fig. 4c indicates the presence of inclusions at an intergranular facet. Ductile features such as dimples are absent, and the observed quasi-cleavage fracture surface does not exhibit the typical hydrogen-induced quasi-cleavage morphology. Similar quasi-cleavage morphology was previously observed at the crack initiation region for the DP-steel during the U-bend test [33]. We cannot determine the role of ferrite in crack initiation and propagation based on fractographic evidence. To this end, we performed site-specific focused ion beam (FIB) lift-out to study the microstructure immediately underneath the fracture surface using transmission Kikuchi diffraction (TKD) and transmission electron microscopy (TEM). These results are presented in detail in section 3.4.

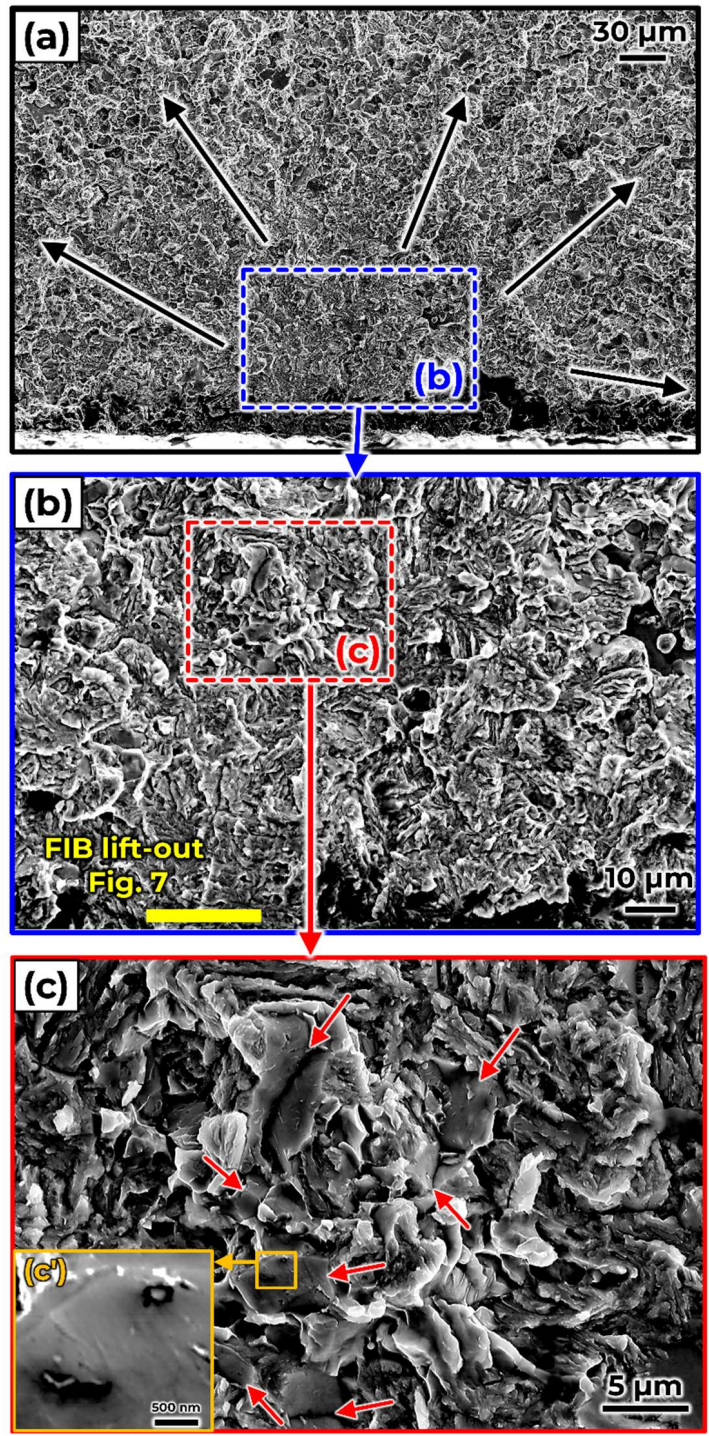

***Figure 4*** *(a) Fractography of the area of crack initiation. Black arrows indicate radial crack propagation. Intergranular fracture facets are highlighted by red arrows in (c). (c') shows the presence of inclusions at an intergranular facet.*

In Fig. 5, we show the fractography of the area adjacent to the crack initiation region (as highlighted in Fig. 3b) at the center of the specimen (along the thickness). We observe the presence of both dimples and quasi-cleavage features in the magnified micrograph in Fig. 5b. It suggests that the local failure mode is a mixed-mode fracture wherein both ductile and brittle fracture modes co-exist.

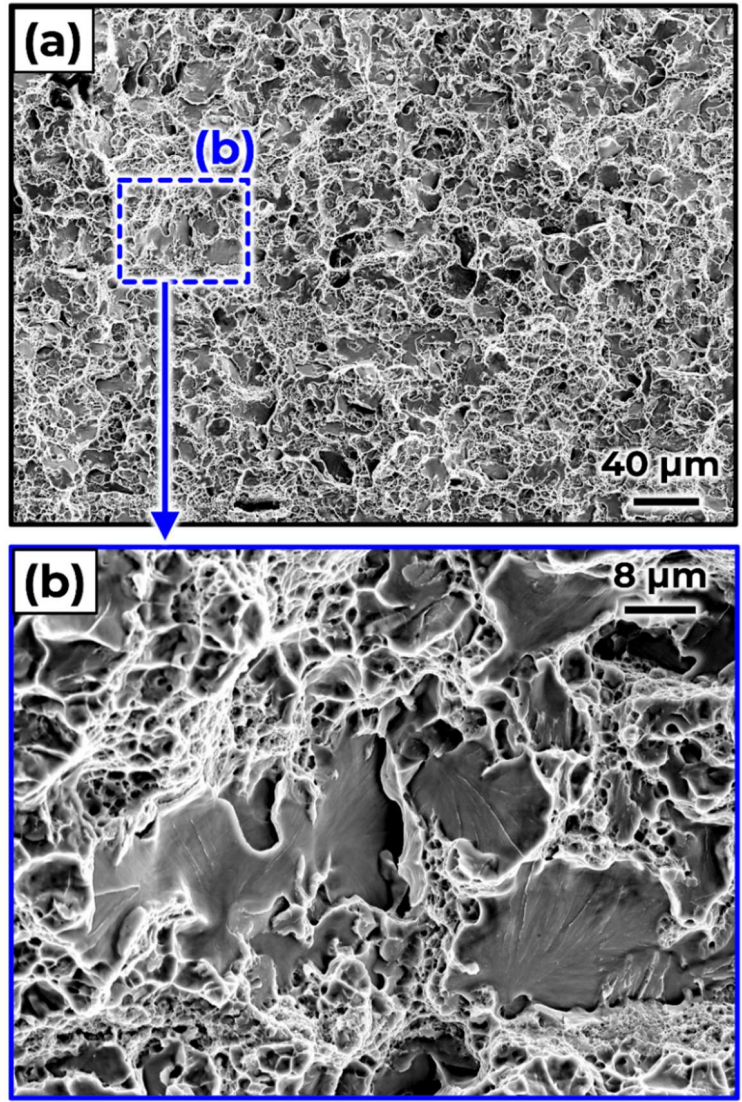

*Figure 5* *Fractography of the area ahead of the region of cracking initiation, as highlighted in Fig. 3(b). A mixture of ductile fracture features of dimples and quasi-cleavage fracture is observed.*

Fig. 6 depicts the fracture surface upon significant crack growth from one end of the tensile specimen to the other; the location is highlighted in Fig. 3a. The magnified fracture surface of the region at the center of the tensile specimen (along the thickness) and the area close to the shear lip at the surface of the specimen are shown in Fig. 6b and 6c, respectively. The location of Fig. 6b and 6c are highlighted in Fig. 6a using blue and red rectangular boxes, respectively. We observe that the fracture surface in Fig. 6b consists predominantly of brittle cleavage and quasi-cleavage features with the presence of occasional voids. On the contrary, in the region adjacent to the shear lip (Fig. 6c), mixed-mode fracture consisting of both brittle and ductile fracture features was observed. At the center of the tensile specimen, two cleavage crack propagation morphologies were observed: linear (Fig. 6d) and zig-zag (Fig. 6e). In the previous study [33] (fracture in DP steel in the absence of hydrogen), linear crack morphology was attributed to {100} ferrite cleavage and zig-zag crack morphology was attributed to transgranular cracking in the lath martensite.

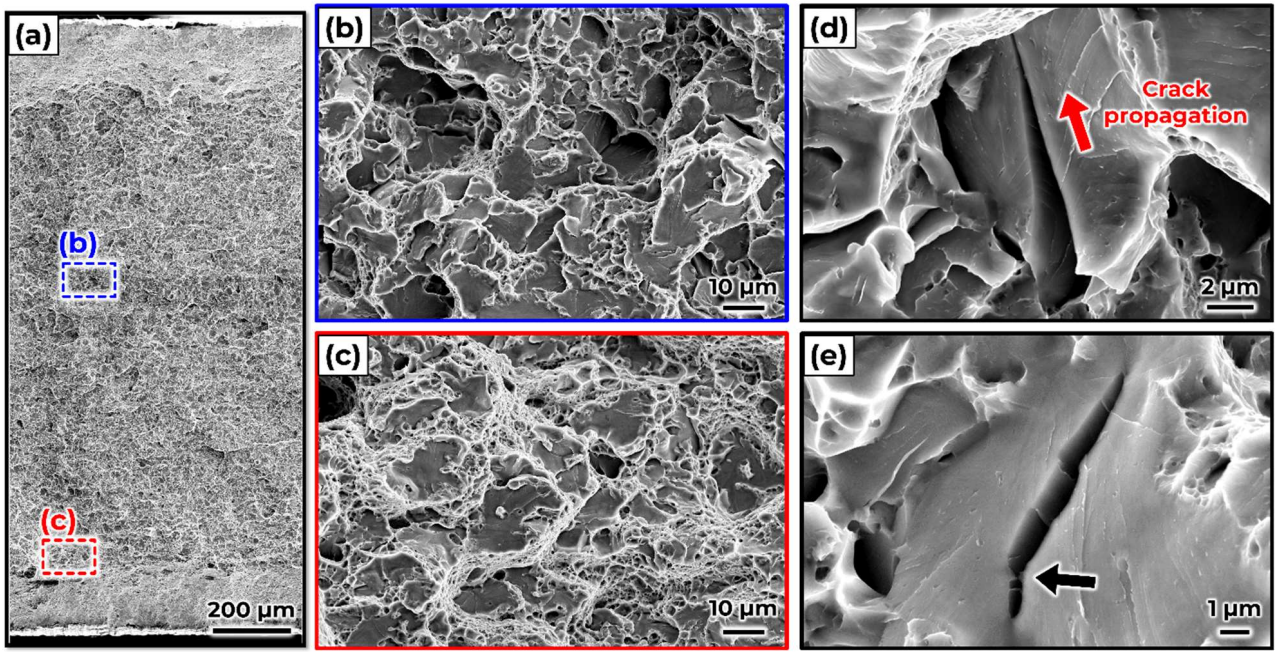

*Figure 6* (a) Fracture surface after significant crack propagation from one end of the tensile specimen to the other (location highlighted in Fig. 3a). **(b)** Predominantly brittle cleavage and quasi-cleavage features were observed at the center of the specimen (indicated with a blue rectangular box in Fig 6a). (c) Mixed-mode fracture consisting of both ductile and brittle fracture features (highlighted with a red rectangular box in Fig. 6a). Two crack propagation morphologies were observed: (d) linear and (e) zig-zag (black arrow).

A hydrogen pre-charged tensile specimen with a strain rate of $10^{-3}$ s$^{-1}$ exhibited fractographic features (shown in supplementary Fig. S1) similar to the strain rate of $10^{-4}$ s$^{-1}$ presented in section 3.3:

1. Surface crack initiation wherein the fracture surface consists of intergranular fracture and quasi-cleavage fracture features.
2. Crack growth initially occurs towards the center from the surface (across the thickness). Subsequently, the crack propagates across the width of the tensile specimen, from one end of the tensile specimen to the other.
3. We examined the fracture surface after significant crack propagation along the width of the specimen. The local fracture at the center of the specimen (along the thickness) is predominantly brittle, and the local fracture mode at the area adjacent to the shear lip was mixed-mode with both brittle fracture features and ductile dimples.

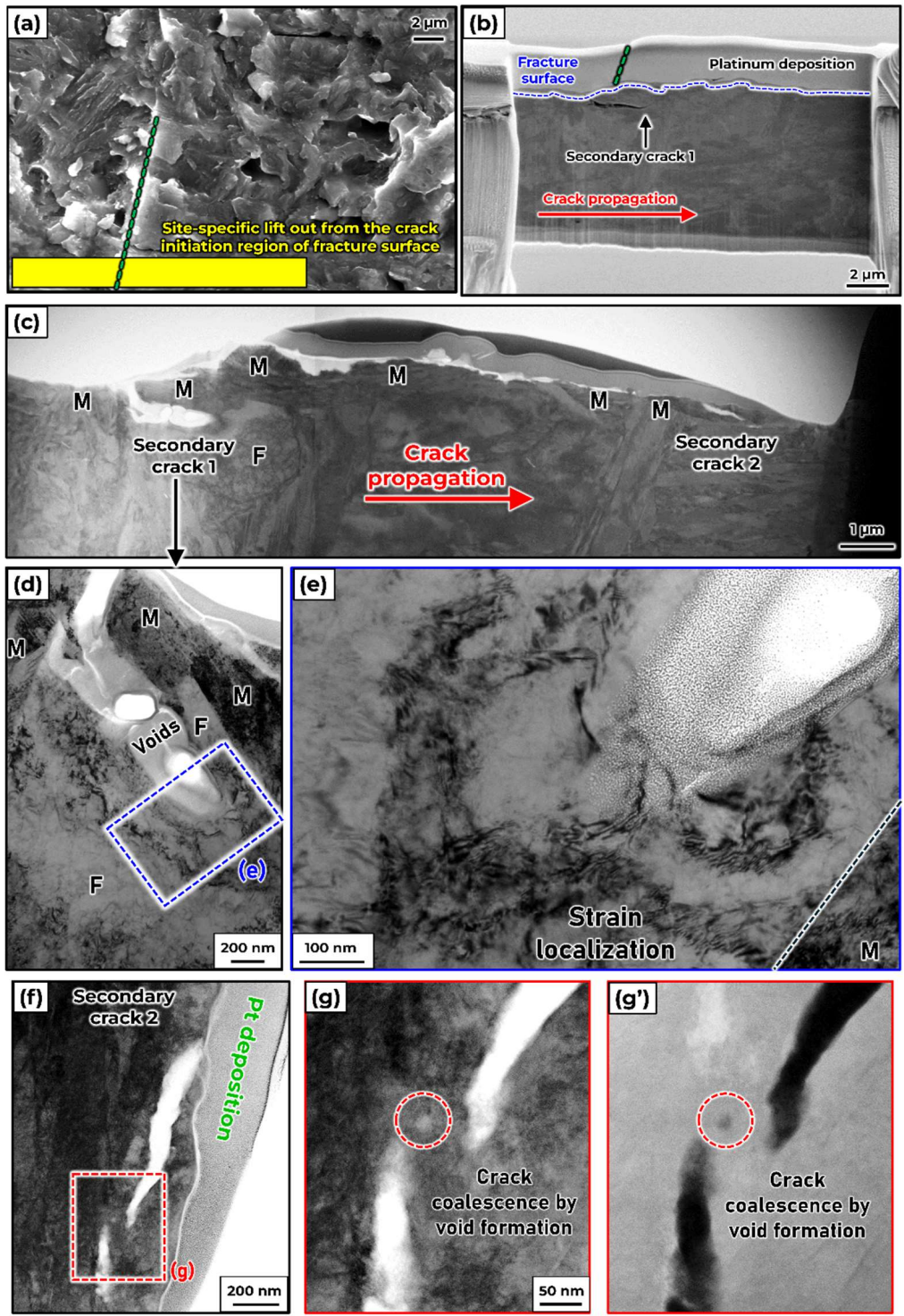

*Figure 7* *(a) Site-specific lift-out of the TEM lamella from the fracture surface at the crack initiation area. (b) TEM lamella before final polishing, indicating the presence of a secondary sub-surface crack. Green dotted line correlates the same feature in (a) and (b). (c) Bright-field image of the lamella after final low-kV polishing. Red arrows in (b) and (c) correspond to the crack propagation direction. M and F in Figs. c-e are abbreviations of martensite and ferrite, respectively. (d) Void formation in ferrite and (e) crack blunting & arrest via strain localization in ferrite ahead of the crack tip. Coalescence of secondary crack 2 observed in (f) via void formation confirmed by high magnification (g) bright-field and (g') annular dark-field images.*

### 3.4 Site-specific transmission electron microscopy (TEM) investigations beneath the fracture surface

Site-specific FIB lift-out at the region of crack initiation is shown in Fig 7a. We observe the presence of a secondary crack in the lift-out underneath the fracture surface, as seen in Fig 7b. Based on the secondary crack, we can infer the crack propagation direction (left to right in Fig. 7b). M and F in Fig. 7c correspond to martensite and ferrite, respectively. In addition to the secondary crack noted in Fig. 7b (crack 1), we observed an additional secondary crack 2, as highlighted in Fig. 7c. Void formation, a typical feature of ductile failure, is evident in Fig. 7d. Additionally, strain localization in ferrite, ahead of the crack tip is observed in Fig. 7e. Figs. 7c-e clarify that ferrite blunts and arrests the crack propagation. Figs. 7f-g' demonstrate the secondary crack coalescence in martensite via void formation.

The lower magnification micrograph in Fig. 8a corresponds to the fracture surface after significant crack propagation (highlighted as the yellow rectangular box in Fig. 3). Figs. 8b-d indicate brittle fracture features at the center of the specimen (along the thickness). It is consistent with the observations presented in section 3.3. We performed site-specific lift-out at the fracture surface shown in Figs. 8c-d. Based on the river pattern at the area of FIB lift-out, the local crack propagation direction was identified (indicated with white arrows in Figs. 8c-d). The local crack pathway is from the center of the tensile specimen towards the surface. It is consistent with the fractography in section 3.3 (Fig. 3a), wherein mesoscopic fracture features indicated that the crack propagation occurs from the center towards the surface of the specimen (along the thickness). However, the local crack pathway differs from the macroscopic crack propagation direction (left to right, Fig. 3a). The presence of a void was observed during the thinning process of the lamella (Fig. 8e). Correlating TKD observations (Figs. 8f-f') with fractography (Figs. 8c-d) reveals that the grain with river pattern is ferrite. The crack propagation direction is out of the plane of the lamella. Trace analysis on the inverse pole figure (IPF) map (Fig. 8f) suggests that ferrite underwent {100} cleavage. The failure mode of the ferrite at the center of the specimen upon significant crack propagation is brittle, unlike at the region of crack initiation (Fig. 7d), where the ferrite blunted the crack and failed in a ductile manner. The kernel average misorientation (KAM) map (Fig. 8f') indicates strain localization in ferrite.

We performed TEM studies to clarify the dislocation structure underneath the {100} cleavage fracture (Figs. 9 a-b) and understand the formation mechanism of river pattern features in ferrite (Figs. 9c-f'). We observe that the dislocation density underneath the fracture surface (red arrow, Fig. 9a) is significantly greater compared to the region distant from the fracture surface (green arrow, Fig. 9a). Hence, we can conclude that the {100} cleavage is accompanied by significant local plasticity adjacent to crack tip. We confirmed that the position of the void observed during the thinning process (Fig. 8e) correlates with the inclusion/precipitate interface (Fig. 9a). Significant localized plasticity is observed adjacent to the inclusion/precipitate. Considering that the inclusion/precipitate is three-dimensional, the void formation observed in Fig. 8e (milled during the thinning process) can be attributed to the localized plasticity surrounding the inclusion/precipitate. Additionally, strain localization was observed between the inclusion/precipitate and the ferrite-martensite phase boundary. This observation is consistent with our previous findings, which showed void formation at MnS inclusions and associated localized plasticity in the condition without hydrogen pre-charging [33]. We observe the formation of micro deformation bands along {110} up to ~2 μm away from the fracture surface (Fig. 9b). Two serrated features (Figs. 9c-f') corresponding to the river-like fracture features in Fig. 8c-d were examined. Serrated feature 1 comprised of {110} and {112} planes, as seen in Fig. 9d. The plane connecting the {110} and {100} planes (red dotted line in Fig. 9d) does not correspond to any low-index plane. Serrated feature 2 only comprised of {110} <100> plane and a connecting plane (red dotted line in Fig. 9f), not correlating any low-index plane. Based on the experimental evidence, mechanisms for the formation of the river pattern features will be discussed in section 4.6.

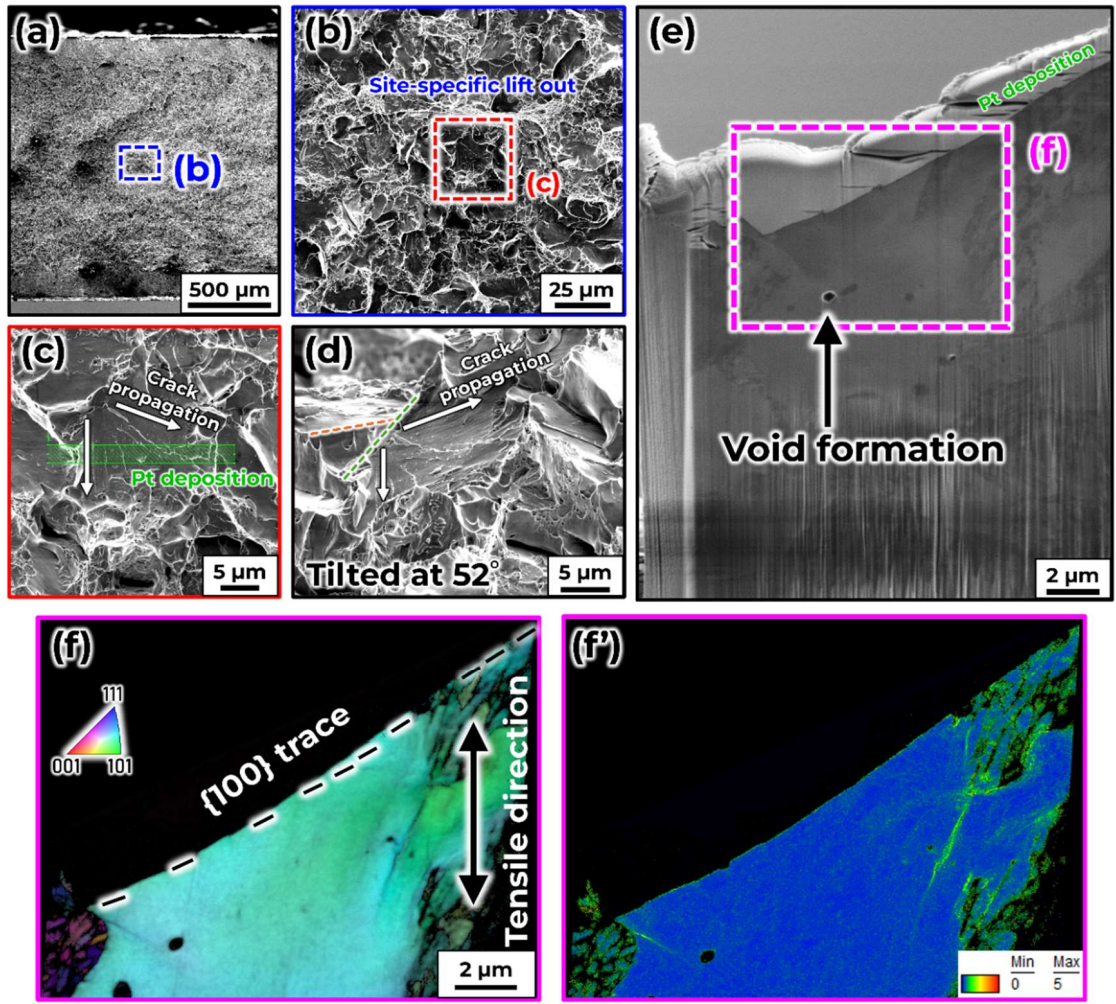

*Figure 8* (a) SEM micrograph indicating the fracture surface after significant crack propagation; the area of (a) is highlighted with a yellow rectangular box in Fig. 3. Higher magnification fractographs of the region of site-specific lift-out are shown in figures (b) and (c). (c) The green shaded region (Pt-deposition) indicates the FIB lift-out area. (d) The fracture surface at a tilt of 52° corresponds to (c). The local crack propagation direction (white arrows) is determined based on the river pattern brittle fracture features. Note that the macroscopic crack propagation direction is from left to right. The green dotted line indicates the trace of the cleavage plane of the grain corresponding to the lift-out. The orange dotted line corresponds to the trace of the cleavage plane of the adjacent grain. (e) Lamella before final polishing, wherein void formation was observed. The magenta dotted rectangular box corresponds to the transmission Kikuchi diffraction (TKD) studies shown in (f-f'). (f) Inverse pole figure (IPF) map and (f') kernel average misorientation (KAM) map. Based on trace analysis, the brittle fracture in ferrite corresponds to {100} cleavage

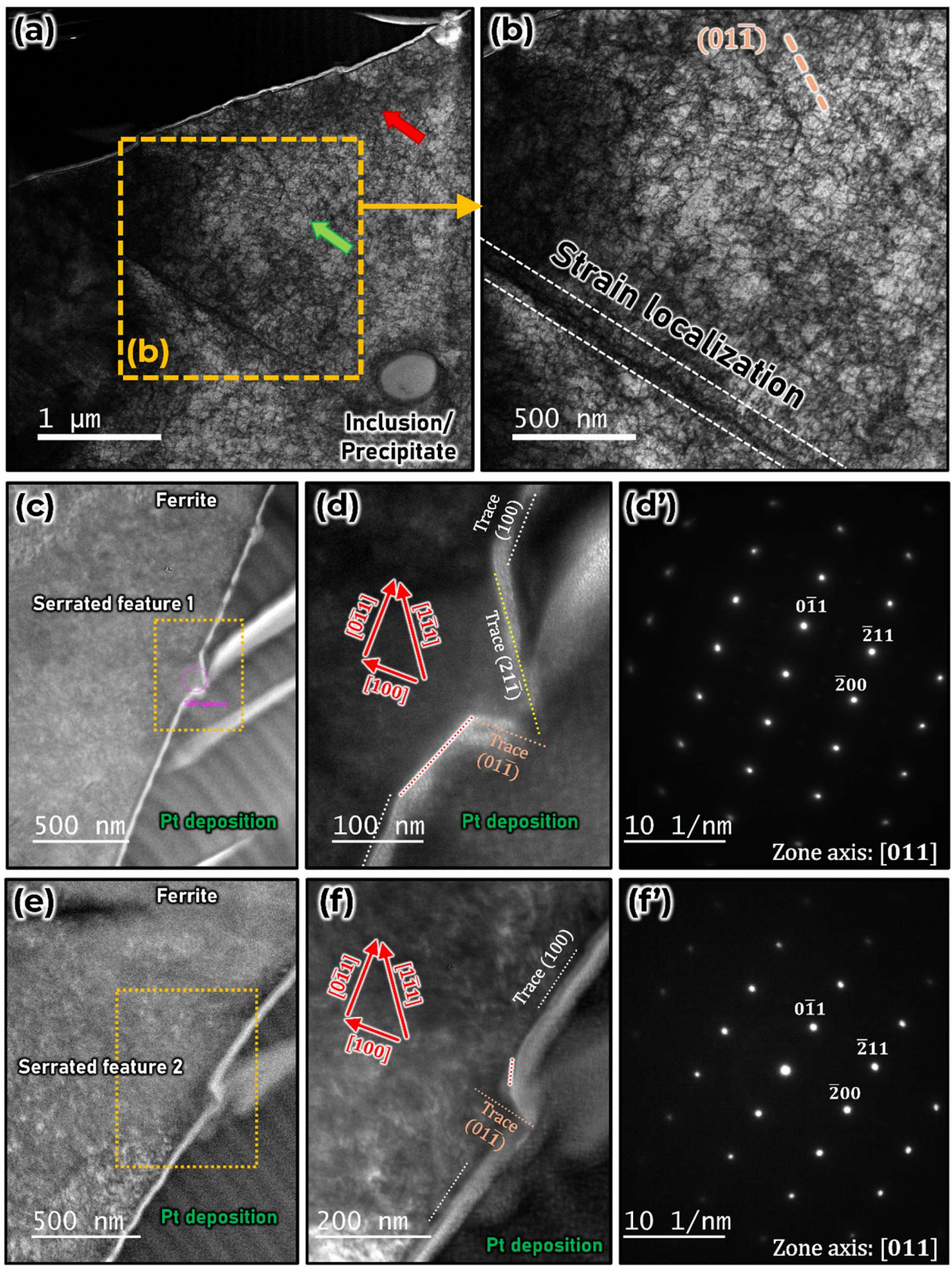

*Figure 9* *(a) Dark-field (DF) electron micrograph revealing the dislocation structure beneath the ferrite fracture surface. The orange dotted box region corresponds to the DF image at a higher magnification in (b). Red and arrows highlight the dislocation density near and away from the fracture surface, respectively. (c-d) Serrated feature 1 in the ferrite fracture surface and the corresponding (d') diffraction pattern. (e-f) Serrated feature 2 and the corresponding (f') diffraction pattern.*

## 3.5 Cross-sectional microstructural analysis of crack initiation and propagation:

Fractography must be correlated with surface and cross-sectional microstructural analysis of crack initiation and propagation to comprehensively understand the failure mechanism(s). Fig. 10 shows representative martensite cracking along the PAGBs (confirmed by the disorientation across the crack in Fig. 10b). Additionally, the KAM map (black arrow in Fig. 10c) indicates strain localization in ferrite ahead of the martensite crack. In other words, ferrite blunts the propagation of hydrogen-assisted cracks formed in martensite.

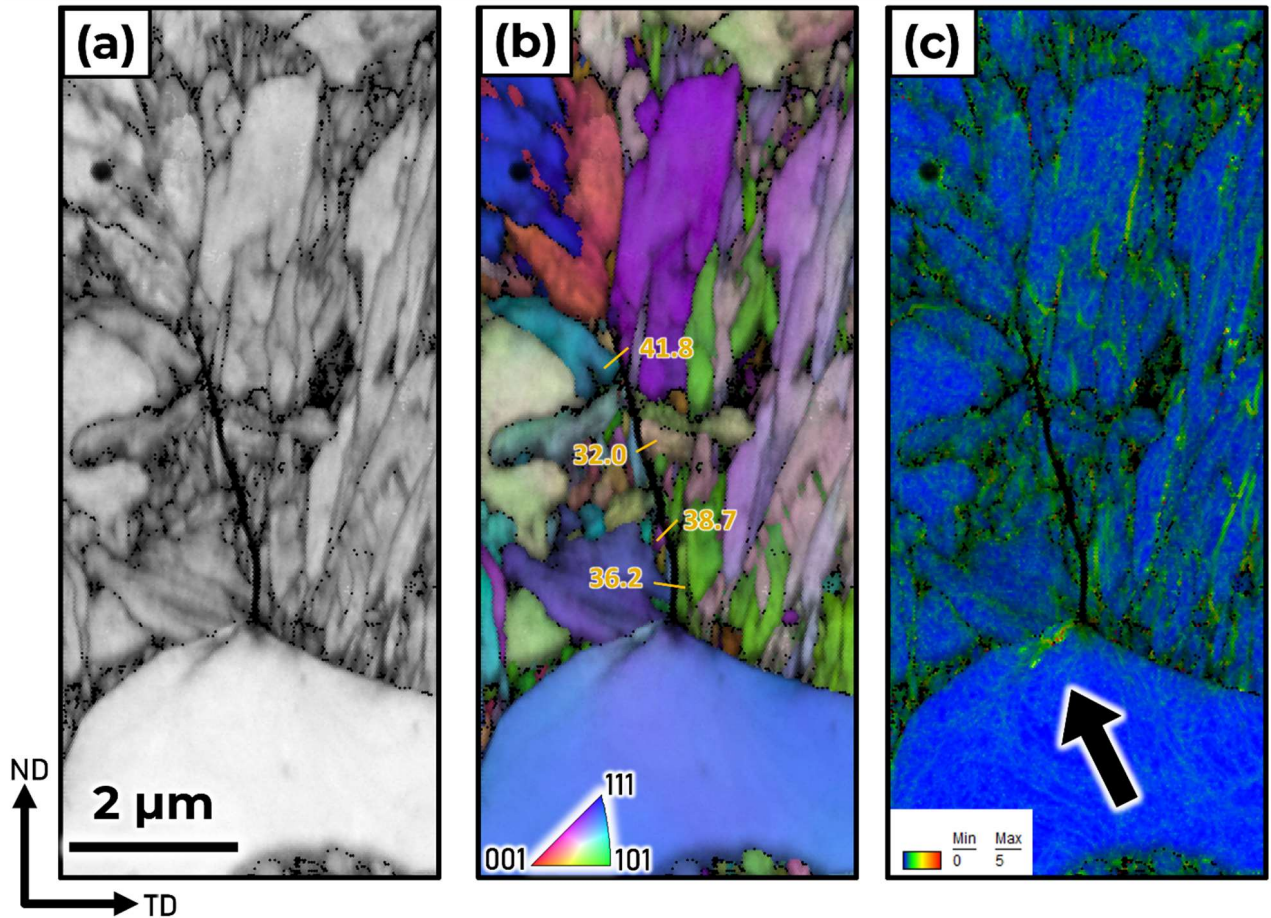

*Figure 10* EBSD analysis of a cross-sectional crack; ND and TD correspond to normal direction and tensile direction, respectively. (a) Image quality (IQ) map indicating martensite cracking and the corresponding (b) IPF map and (c) KAM map. Analysis of the disorientations (Fig. b) across the crack (highlighted in yellow) confirms that the crack propagates along a PAGB. The black arrow in the KAM maps indicates strain localization in ferrite ahead of the PAGB crack tip.

Multiple surface and edge cracks were observed on the tensile specimen after fracture (supplementary Fig. S2a). Fig. 11 shows a microstructural analysis of an edge crack, wherein the crack tip was blunted by ferrite, evidenced by the strain localization in KAM (Fig. 11 b') and ECC micrographs (Fig. 11 c). Crack blunting by ferrite was also observed during the crack propagation, indicated by white arrows in Figs. 11 d-d'. Based on the crack tip opening displacement, it can be concluded that a fresh crack formed at the ferrite-martensite interface ahead of the arrested crack. The martensite crack paths predominantly correspond to {100} or {110} cleavage cracking. We used two-dimensional EBSD trace analysis to index the crack path. Hence, the trace of the cleavage plane could correspond to either of {100} or {110} planes (for instance, when the crack path is along <001> direction) and has been labeled as 'Either'. Occasionally, transgranular cracking that does not correspond to either of {100} and {110} planes was observed.

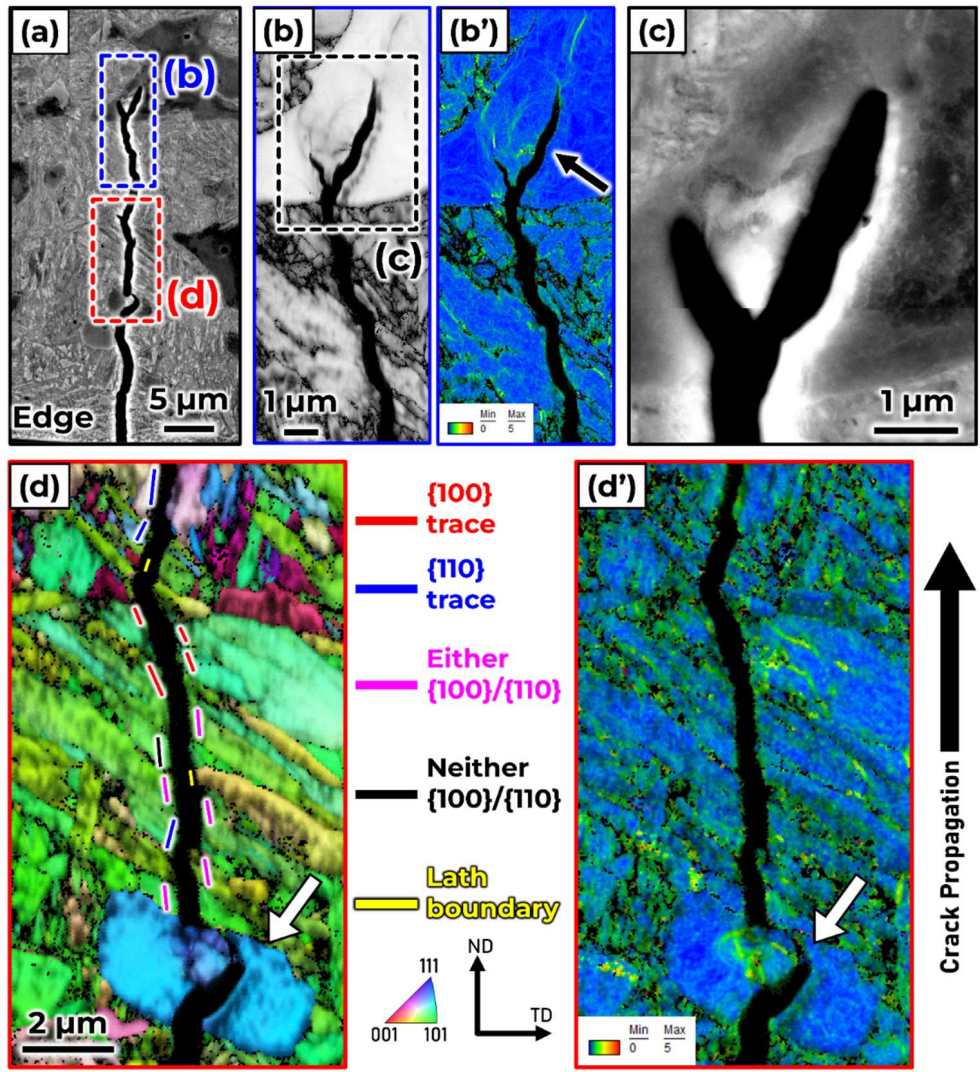

*Figure 11* (a) Secondary electron image of an edge crack. (b) IQ map and the corresponding (b') KAM map of the crack tip (blue rectangular box in (a)). The black arrow highlights the strain localization in ferrite ahead of the blunted crack. (c) ECC micrograph of the blunted crack tip (highlighted in (b) with a black rectangular box). (d) IPF map (with IQ map overlay) showing the crack pathways in martensite based on trace analysis and the corresponding (d') KAM map. ND and TD correspond to normal and tensile directions, respectively. The white arrow shows crack blunting in ferrite and a fresh crack formation at the ferrite-martensite interface.

## 4 Discussion

Hydrogen embrittlement in DP steels with tensile strength ≤ 1.2 GPa has been extensively investigated [21,36–39]. Recently, we investigated the hydrogen-induced delayed fracture in the 1.5 GPa DP steel via U-bend testing [5,40]. However, an understanding of the hydrogen embrittlement mechanism in tensile loading conditions for the 1.5 GPa DP steel is lacking. From a microstructure design perspective, understanding the role of soft ferrite in the hydrogen embrittlement mechanism is essential for developing novel high-strength DP steels with improved resistance to hydrogen embrittlement, which is the aim of the current work.

### 4.1 Limitations of fractography in understanding hydrogen embrittlement

We previously observed that, in the absence of hydrogen, the tensile sample showed nearly complete brittle fracture features after failure with ductility of 5.21 % [33]. Upon hydrogen pre-charging, the tensile specimens exhibited 1.13% and 0.57% ductility at strain rates of $10^{-3}$ $s^{-1}$ and $10^{-4}$ $s^{-1}$, respectively. While the uncharged sample exhibited predominant brittle fracture [33], pre-hydrogen-charged specimens showed mixed-mode fracture features (ductile + brittle), as shown in section 3.3. In other words, despite the ductile fracture features observed in fractography, the tensile specimens exhibit significant embrittlement in the presence of hydrogen. Brittle fracture (such as hydrogen embrittlement) is often correlated with the extent of brittle fracture features (area fraction) obtained from fractographic analysis; the current work shows the limitations of such an approach.

### 4.2 Crack initiation and small crack growth

Hydrostatic stress gradients provide a driving force for the diffusion of lattice hydrogen [41,42]. In a DP steel, Depover et al. [43] showed that hydrogen localizes ahead of the notch tip at the area of maximal hydrostatic stress concentration, resulting in crack initiation. Given the stress partitioning in the DP steel [33,44], under tensile loading, martensite would be subject to higher stress compared to ferrite, thus resulting in hydrogen partitioning. The surface roughness (supplementary Fig. S2a) can accentuate the hydrostatic stress, causing crack initiation adjacent to the surface of the tensile specimen.

EBSD studies of the cross-sectional micro-crack (Fig. 10) indicate that crack initiation occurs at the PAGBs. The intergranular fracture facets in Fig. 4c provide further evidence of PAGB crack initiation. As noted in the previous study, PAGBs act as crack initiation sites even in the absence of hydrogen [33]. In the presence of hydrogen, PAGB cracking occurs within the elastic limit (section 3.2), which can be attributed to the HEDE of the PAGB [45]. Furthermore, hydrogen accumulates at the PAGBs during deformation at a lower strain rate [46] due to microscopic stress-induced diffusion during loading [47]. The already brittle PAGBs (given they crack in the uncharged specimen [33]) would readily undergo enhanced decohesion due to hydrogen accumulation during tensile loading. Notably, no decohesion of the ferrite-martensite interface was observed either in the hydrogen-charged or uncharged specimen [33]. The inset Fig. 4c' reveals sub-micron inclusions (likely MnS) at the intergranular facet. Void formation at MnS inclusions in the same material but without hydrogen has been previously established [33]. Hence, the HEDE of the PAGB is likely to be further aggravated by MnS inclusions at the PAGB.

The crack formation increases the local hydrostatic stress at the PAGB, enhancing the stress-induced diffusion of hydrogen to the PAGB. A plausible subsequent process of small crack growth has generally been argued that the repetition of hydrogen accumulation and cracking results in quasi-cleavage fracture along {110} [48]. However, the fracture surface morphology at the crack initiation region (Section 3.3 and Fig. 4) starkly differs from the typical hydrogen-related quasi-cleavage morphology. TEM studies (Section 3.4 and Fig. 7) revealed that the fracture morphology corresponds to quasi-cleavage in martensite. While quasi-cleavage fracture is evident, we could not experimentally identify the local cleavage paths. The strain localization ahead of secondary crack 1, accompanied by void formation, as observed in Figs. 7b-e demonstrates that ferrite blunts the crack at the crack initiation region. The ineffective prevention of crack growth can be attributed to a lower ferrite fraction (25%) since it is insufficient to disrupt the network of martensite along which the crack propagates.

Furthermore, TEM and TKD results (Figs. 8 and 9) of site-specific lift-out from fracture surface upon substantial crack growth (at the center of the specimen) reveal that ferrite undergoes {100} cleavage. However, it is unlikely that the ferrite {100} cleavage was directly assisted by hydrogen. Okada et al. [48] previously reported that at a strain of 3% while the hydrogen-charged specimen exhibited a tangled dislocation morphology, the dislocations were linear in the uncharged specimen. In the current work, we neither observe {110} quasi-cleavage nor tangled dislocation morphology in the ferrite. While the velocity of brittle crack propagation in steel was estimated to be ~1000 m s$^{-1}$ [49], the hydrogen diffusion distance at room temperature in 1 s is 10$^{-4}$–10$^{-6}$ m [50,51]. Consequently, it is improbable for the hydrogen to diffuse ahead of the crack tip during brittle fracture; hence, crack propagation is unlikely to be directly assisted by hydrogen. Hence, the crack growth in the present case is unconventional. We discuss this unconventional crack growth behavior based on the Hydrogen-induced fast fracture model.

### 4.3 Hydrogen-induced fast fracture model

We briefly summarize the hydrogen-induced fast fracture (HIFF) model proposed by Shishvan et al. [52].

<u>Stage A: Hydrogen-assisted crack growth</u>

(i) Tensile loading results in the debonding of inclusions or expansion of pre-existing flaws, resulting in the formation and growth of cavities. These cavities are filled by hydrogen diffusing from the surrounding matrix.

(ii) During tensile loading, stress concentrations are generated at the nano cracks on the cavity surface. Hydrogen-assisted cleavage initiation was ascribed to the nano-scale hydride (FeH$_y$, y ≈ 1-3) formation ahead of the crack tip.

(iii) The cavity supplies hydrogen for hydride formation, sustaining hydrogen-assisted cleavage crack speeds exceeding 100 m s$^{-1}$. The high crack velocity prevents effective crack blunting via dislocation emission. It enables cleavage cracking until the cavity hydrogen is depleted, marking the end of stage A.

We note that the formation of nano-scale hydride has never been experimentally observed in ferrous alloys. We proposed an alternative mechanism for stage-A crack initiation in section 4.2 wherein the cracking process in the extremely vulnerable PAGB (due to HEDE) would not require the assistance of hydride. Furthermore, the hydrogen diffusion to the crack PAGB crack tip ensures the hydrogen supply for intergranular crack growth (stage-A propagation) to sustain a crack velocity > $v_{crit}$ for martensite. The fast crack growth limits dislocation emission [52], thus maintaining the brittle crack propagation within the martensite. Schematic of the crack initiation mechanism is presented in Fig. 12.

Stage B: Crack growth in the absence of hydrogen

(iv) The crack tip plasticity is restricted by the high-strain rate at the crack tip caused by the hydrogen-assisted fast crack growth. Consequently, as long as the externally applied stress is sufficiently high, brittle crack propagation persists over long distances even without hydrogen supply from the cavity. This fast crack growth culminates in the macroscopic brittle fracture of the specimen.

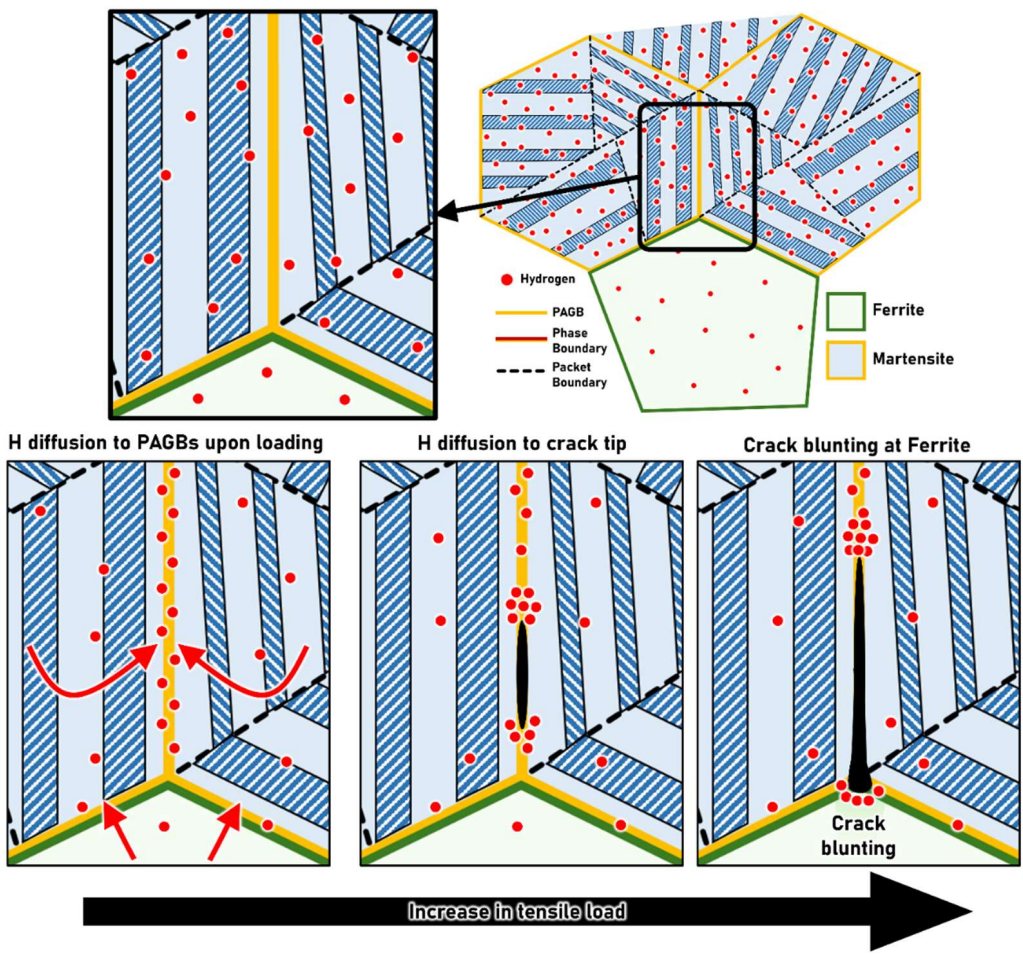

**Figure 12** *Schematic of crack initiation via hydrogen-enhanced decohesion at the PAGB*

### 4.4 Crack propagation

Ahead of the region of crack initiation (Fig. 5), mixed-mode fracture consisting of both ductile and brittle features was observed. The ductile features can be attributed to ferrite since ferrite arrested cracking at the crack initiation region. Upon significant crack propagation, the fracture surface morphology exhibits predominant brittle fracture at the center of the specimen (Fig. 6b) and mixed-mode fracture (ductile + brittle) close to the shear lip at the surface of the specimen (Fig. 6c).

Note that in the pre-hydrogen-charged tensile specimen, the crack initially propagates from the surface towards the center and subsequently undergoes crack deflection (propagating from one end to another). Initially, for a short crack length, the local strain rate at the crack tip is insufficient to overcome the dislocation emission in ductile ferrite, while martensite undergoes brittle fracture. This is evidenced by fractography consisting of ductile fracture features adjacent to the region of crack initiation (Fig. 5). In other words, while the crack propagation in martensite is brittle, ferrite failure is not brittle.

Shishvan et al. [52] explained that the hydrogen-assisted fast crack growth would induce a high strain rate at the crack tip. In turn, the high strain rate limits plasticity at the crack tip, ensuring brittle crack propagation even in the absence of hydrogen supply to the crack tip, eventually resulting in macroscopically brittle fracture (stage-B) [52]. Shishvan et al. [52] clarified that complete suppression of dislocation emission is not necessary; as long as the crack velocity > $V_{crit}$ (5.4 ms$^{-1}$ for BCC Fe), the brittle crack propagation would be able to overcome crack tip blunting by dislocation emission. Since for a given external load, the stress intensity factor (K) ∝ (crack length)$^{0.5}$, as the crack length increases, K and crack velocity increase. Based on the stage-B of fast fracture proposed by Shishvan et al. [52], when the crack velocity > $v_{crit}$ for ferrite, brittle {100} ferrite cleavage is expected. Once a critical crack velocity is reached (due to the high strain rate at the crack tip), despite the absence of hydrogen supply to the crack tip, ferrite can undergo {100} cleavage, overcoming the dislocation emission. The enhanced dislocation density observed adjacent to the {100} cleavage in Fig. 9a correlates to dislocation emission ahead of {100} cleavage. Our previous report of brittle {100} cleavage in ferrite in the absence of hydrogen [33] further supports this hypothesis. In the uncharged tensile specimen, {100} ferrite cleavage was attributed to the high strain rate at the crack tip caused by stress partitioning between ferrite and martensite [33]. In the uncharged DP steel, at a macroscopic strain of 5%, the stress in martensite can be ~1700 MPa (~800 MPa greater than ferrite) [33]. In the hydrogen pre-charged specimen, the brittle fracture occurs within the elastic limit at a much lower stress of 900 MPa (section 3.2). Hence, the high strain rate at the crack tip cannot be due to stress partitioning. Instead, the high strain rate at the crack tip is caused by the HIFF in the pre-charged tensile specimen.

The presence of ductile features adjacent to the shear lip (Fig. 6c) can be attributed to the change in fracture mode from plane strain to plane stress as the crack propagates from the center to the surface. The crack initially reaches the center of the thickness and propagates from one end of the tensile specimen to the other (along the width) at the center (Fig. 3a). Once the crack propagates through the center, the crack needs to grow along the thickness, from the center to the surface (Fig. 3a). In other words, there is a change in the local crack propagation direction, as the crack grows along the thickness, from the center to the surface. As a result, when the crack propagates from the center to the surface along the thickness, the effective crack length for mode I opening reduces (Fig. S3), reducing the stress intensity factor at the crack tip. Consequently, the crack velocity would drop below the critical threshold required to suppress dislocation emission in ferrite, leading to ductile fracture of ferrite.

Fig. 11 shows EBSD analysis of an edge crack of length ~ 35 µm. Due to the short crack length (and ferrite blunting), the crack growth is expected to be gradual and controlled. We observe that the ferrite arrests the crack, and a fresh crack forms at the ferrite-martensite interface ahead of the arrested crack. This behavior is identical to the observation in the U-bend test [5]. The transgranular (trans lath) martensite cracking observed in Fig. 11d (section 3.5) occurs predominantly via {110} and {100} cleavage cracking. Hydrogen-induced {110} cleavage in martensitic steels is well documented in the literature [53–55]. Furthermore, in the U-bend test, the trans lath cracking within martensite occurred via {110} cracking [5]. However, the DP steel exhibited {110} martensite cleavage even in the absence of hydrogen [33]. Therefore, based on the {100} + {110} cleavage cracking (Fig. 11d), we cannot confirm if the transgranular cracking in martensite is hydrogen-assisted. Secondary crack 2 (Figs. 7f-g') provides evidence of crack coalescence via void formation. Furthermore, it also illustrates discontinuous cracking with the formation of a fresh crack ahead of a crack tip and subsequent coalescence. Similar discontinuous cracking was observed at an edge crack, as shown in Supplementary Figs. S2b-c. The coalescence of discontinuous cracks is likely to be plasticity-driven, and the formation of specific crystallographic cleavage planes is unlikely.

Table 2 *Summary of crack initiation, arrest, and propagation in ferrite and martensite for 1.5 GPa DP steel.*

| | Martensite | | | | Ferrite | | | |
|---|---|---|---|---|---|---|---|---|
| | Uncharged | With Hydrogen | | | Uncharged | With Hydrogen | | |
| | Tensile [33] | Tensile (current work) | | U-Bend test [5] | Tensile [33] | Tensile (current work) | | U-Bend test [5] |
| | | Crack initiation region | Significant crack growth | | | Crack initiation region | Significant crack growth | |
| Crack Initiation | PAGB | | | | - | | | |
| Crack arrest | X | X | X | X | X | ✓ | X | ✓ |
| Crack Propagation | {110} + others | QC {110} + {100} | - | {110} | {100} | Crack blunting | {100} | Crack blunting |

Table 2 summarizes the crack initiation, arrest, and propagation in the DP steel for the tensile tests (with and without hydrogen) and the U-bend test for hydrogen-induced delayed fracture. The crack propagation behavior ahead of crack initiation in the pre-hydrogen-charged tensile specimen resembles observations in the U-bend test, wherein ferrite arrested the crack and martensite underwent hydrogen-induced quasi-cleavage fracture [5]. In the case of the U-bend test [5], the bolt-tightened specimen is initially charged at a lower current density of 1 A m$^{-2}$. If no fracture was observed after 12 hours, the current density was increased to 10 A m$^{-2}$ [5]. This ensures that the hydrogen uptake would be gradual for the U-bend test, contrary to the hydrogen pre-charged tensile test. Due to a gradual hydrogen supply, the crack velocity remains < $v_{crit}$ in the U-bent test specimen.

Based on the HIFF model, Shishvan et al. [56] proposed that the strain rate sensitivity is governed by the hydrogen desorption rate from the cavities (stage A). In Section 4.3, we explained that the mechanism of cleavage initiation via hydride formation at cavities is unlikely. We proposed a more plausible alternative based on the HEDE of the PAGB. Therefore, the strain rate sensitivity can be attributed to the kinetics of hydrogen diffusion to the PAGB during tensile loading [46,47,57]. Given that the fractographic behavior is identical for the strain rates of $10^{-3}$ s$^{-1}$ and $10^{-4}$ s$^{-1}$, it is reasonable to infer a consistent hydrogen embrittlement mechanism for both strain rates.

### 4.5 Formation of river pattern brittle features on the {100} ferrite fracture surface

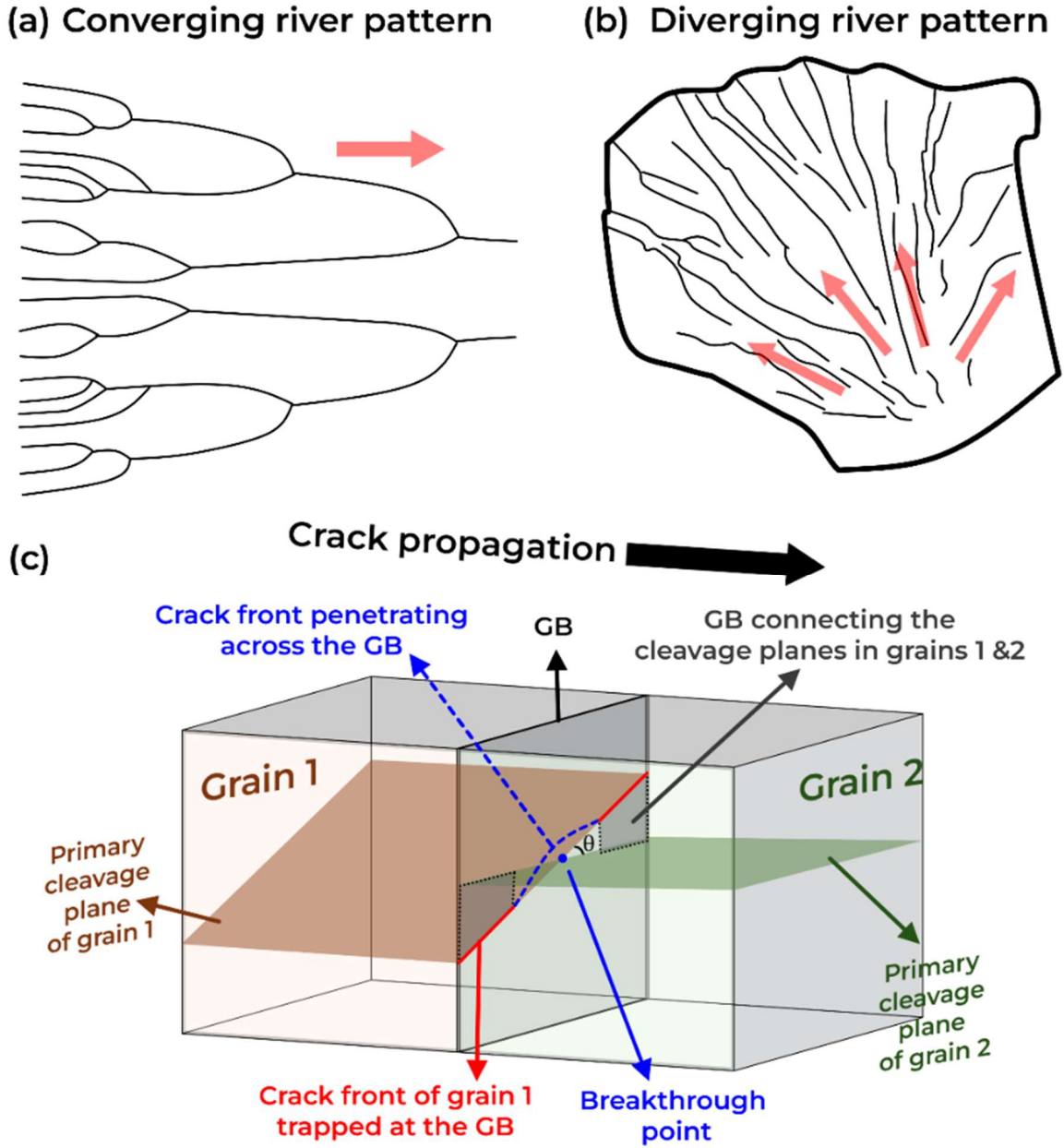

**Figure 13** *Schematics of (a) converging and (b) diverging river pattern. Red arrows indicate the crack propagation direction. (c) Mechanism of formation of diverging river pattern: schematic of a cleavage crack propagation around the breakthrough point across a high angle grain boundary (based on Qiao et al. [58]).*

Firstly, we would like to distinguish between two kinds of river patterns: (i) converging and (ii) diverging (shown in Figs. 13 a-b). In the current work, the observed river pattern (Fig. 8c) is diverging. The schematic in Fig. 13c illustrates the mechanism of formation of the diverging river pattern. As the crack propagates from grain 1 to grain 2, the primary cleavage planes of the two grains are oriented at an angle of θ with each other (Fig. 13c). Such an inclination angle between the cleavage planes is observed in Fig. 8d. Due to this inclination between the cleavage planes, when the crack propagates from grain 1 to 2, there is a local breakthrough point. The crack propagation occurs locally at the breakthrough window (region adjacent to the breakthrough point), and subsequently, the cleavage front grows in grain 2 [58]. In the presence of only one breakthrough point, the cleavage front grows radially from this breakthrough point, and the river pattern features also form radially, thus resulting in a diverging river pattern. In Figs. 8c-d, we observe that the river pattern features are absent at the origin of {100} cleavage in the ferrite grain (Figs. 8c-d), and as the crack propagates, a radially diverging river pattern is observed.

The formation of hydrogen-related quasi-cleavage serrated brittle fracture fractures was previously discussed in ferrite [48,59,60] and martensite [54,61]. However, as discussed in section 4.4, the ferrite {100} brittle cleavage in the pre-hydrogen-charged tensile specimen is not directly assisted by hydrogen. Despite the absence of hydrogen supply to the crack tip, {100} ferrite cleavage fracture occurs because of hydrogen-induced fast crack growth (v > $v_{crit}$) that overcomes crack tip blunting (limited dislocation emission). Given that {100} cleavage is not directly assisted by hydrogen, hydrogen-assisted plasticity mechanisms [59] can be ruled out while understanding the formation of river pattern features. Classically, the formation of river pattern features has been attributed to screw dislocations [62,63]. However, the formation of serrated features with a step size of ~100 nm (Figs 9c-f) cannot be attributed to screw dislocations alone. Alternatively, deflection [54] of the fast fracture crack emerges as the potential serrated feature(s) formation mechanism. Based on the observation of serrated features along {110} and {121} (Fig. 9b), crack deflection can be attributed to two mechanisms:

(i) Twin formation and subsequent twin boundary cracking was reported ahead of the crack tip in ferritic steels [65,66]. Hence, the {121} serrated feature can be attributed to the deflection of the crack along the deformation twins.

(ii) {100} cleavage is typically observed in BCC Fe despite {110} having lower surface energy because dislocation emission from the {100} crack front does not occur easily (compared to {110}) [64]. In Fig. 9b, we observe that the formation of micro deformation bands along {110} can occur over a distance of ~2 μm away from the crack. {110} micro deformation banding occurs ahead of the crack tip before the crack propagates, and the crack tip reaches this {110} micro deformation band. Such a prior micro deformation banding along {110} at the crack tip can alter the micro-stress field and the associated dislocation emission capability. Subsequently, given the low surface energy, local deflection of the crack along the {110} micro deformation band (Fig. 9b) can result in {110} cracking. While we propose a plausible mechanism for the formation of serrated features along {110}, further theoretical validation (such as molecular dynamics simulations) of this hypothesis is required.

The red dotted lines in Figs. 9d and 9f are likely to be coalescence features, given that they do not correspond to any low-index plane. Schematic in Fig. 14 summarizes the formation of serrated river pattern features on {100} cleavage.

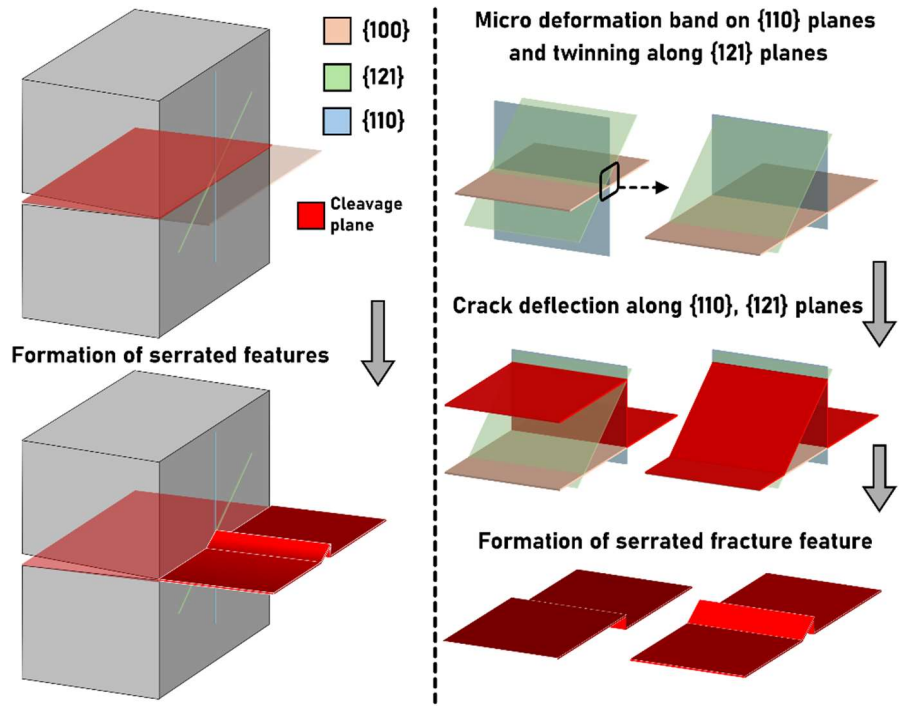

**Figure 14** *Schematic of the formation of serrated features on the {100} cleavage plane.*

### 4.6 Microstructure design strategies to leverage the crack-arresting ability of ductile ferrite

Based on the findings of the present work, we propose strategies to mitigate hydrogen embrittlement in high-strength DP steels by leveraging the crack-arresting ability of soft ferrite.

1. Ferrite helps arrest the cracks during their initiation. Thus, the inability of ferrite to prevent HIFF can be ascribed to its low phase fraction which fails to disrupt the continuous hard martensite network, thereby allowing crack propagation within the martensite. Increasing the ferrite fraction to disrupt the continuity of martensite, however, is expected to reduce the UTS. We can resolve this by designing the ferrite microstructure to consist of fine carbide precipitates. Pelligra et al. [67] have demonstrated that UTS of 1.5 GPa can be achieved with a 40% martensite fraction via vanadium microalloying. The presence of carbides can additionally alleviate the stress partitioning (and thereby hydrogen partitioning) between ferrite and martensite. Since carbides can trap diffusible hydrogen [11], it would activate an additional mechanism to mitigate hydrogen embrittlement.

2. A counterintuitive strategy is to engineer microstructures with a large volume fraction of sub-micron MnS inclusions. Kim et al. [68] have shown that fine or Ti(C,N)–MnS complex inclusions trap diffusible hydrogen effectively in a 2 GPa martensitic steel. The S or S–Ti added steels exhibited enhanced resistance to hydrogen embrittlement compared to the S-free specimens, provided the inclusions were not coarse [68]. We previously observed that MnS promoted plasticity (and strain localization) in ferrite [33]. The resulting strain hardening could provide greater resistance to fast fracture by reducing crack velocity below the threshold.

3. Alternatively, the Cu present in steels recycled from scrap [69] can be used to produce nano-scale Cu precipitates to simultaneously induce precipitation strengthening [70] and trap diffusible hydrogen [12].

## 5    Conclusions

This work provides the first insight into the effect of introducing ductile ferrite as a microstructure design strategy to mitigate hydrogen embrittlement in high-strength steel with tensile strength ≥ 1.5 GPa. However, despite the introduction of 25% ferrite, in the presence of hydrogen, the martensite-ferrite dual phase (DP) steel failed within the elastic limit at 900 MPa upon tensile loading at a strain rate of $10^{-4}$ $s^{-1}$. Site-specific transmission electron microscopy (TEM) revealed that ferrite arrests cracking during crack initiation and early stages of crack growth. However, upon significant crack propagation, {100} ferrite cleavage cracking was observed. This hydrogen embrittlement behavior was explained based on the hydrogen-induced fast fracture (HIFF) mechanism. Crack initiation occurs at prior austenite grain boundaries (PAGBs) due to hydrogen-enhanced decohesion facilitated by hydrogen diffusion during loading. The hydrostatic stress at the crack tip provides further driving force for hydrogen diffusion to the crack front, promoting crack growth at a high velocity. Given the lower ferrite phase fraction, crack propagation within the network of brittle martensite could not be suppressed. Once the crack velocity exceeds a critical threshold ($V_{crit}$), it effectively overcomes blunting via dislocation emission even after the hydrogen supply to the crack tip is depleted. Thus, the fast crack growth explains {100} ferrite cleavage cracking. Additionally, based on TEM observations, we proposed that the crack deflection along {110} micro deformation band and {121} twin result in the formation of river pattern features on the {100} ferrite cleavage surface.

The present study highlights the importance of considering the HIFF mechanism while interpreting brittle fracture in pre-charged high-strength steels, given that hydrogen is unlikely to directly facilitate brittle crack propagation once the critical velocity is attained.

## 6    Acknowledgements:

The authors are grateful to Dr. Makoto Nagasako of the Analytical Research Core for Advanced Materials, Institute for Materials Research, Tohoku University, and Dr. Kosei Kobayashi of the Department of Materials Science and Engineering, Tohoku University, for their valuable assistance with the TEM experiments. R.S. Varanasi thanks Ms. Kaori Sato for support with the FIB facility at the Tohoku University Materials Solution Center (MaSC). Authors gratefully acknowledge the funding received from the project, JPNP14014, commissioned by the New Energy and Industrial Technology Development Organization (NEDO).  R.S. Varanasi is grateful for the postdoctoral fellowship from the Japan Society for the Promotion of Science (Grant number: 24KF0007).

# Hydrogen-induced fast fracture in a 1.5 GPa dual-phase steel

*Rama Srinivas Varanasi[1,2], #Motomichi Koyama[1], Shuya Chiba[1,3], Saya Ajito[1], Eiji Akiyama[1]

[1] *Institute for Materials Research, Tohoku University, 2-1-1 Katahira, Aoba-ku, Sendai, 980-8577, Japan*
[2] *Department of Materials Science & Metallurgical Engineering, Indian Institute of Technology, Hyderabad-502285, India*
[3] *Graduate School of Engineering, Tohoku University, 6-6-01-2 Aramaki Aza Aoba, Aoba-ku, Sendai, Miyagi, 980-8579 Japan*

**Corresponding authors:**

*Rama Srinivas Varanasi: varanasiramasrinivas@gmail.com

#Motomichi Koyama: motomichi.koyama.c5@tohoku.ac.jp

## Supplementary

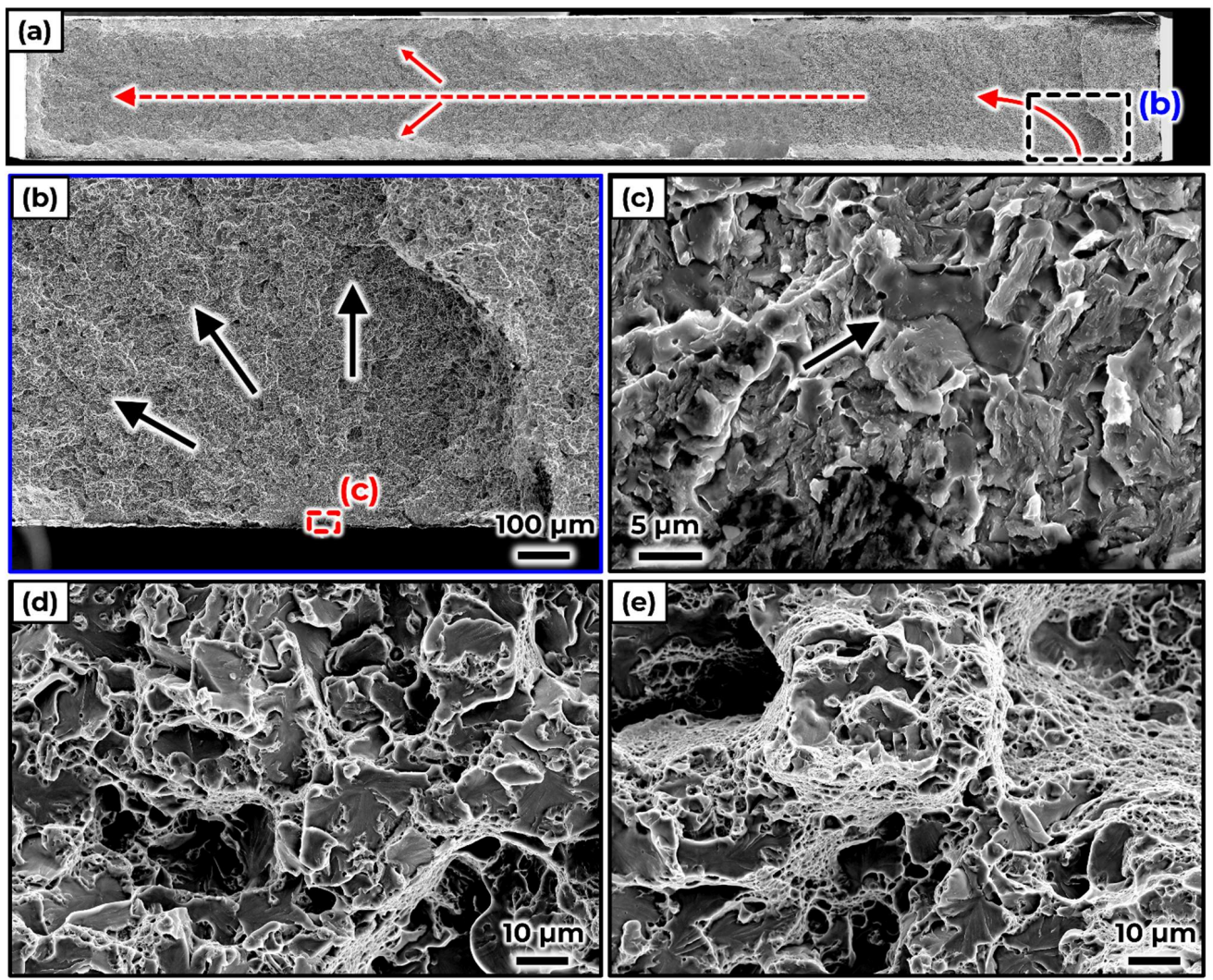

**Figure S1** *(a) Fracture surface of the DP steel tensile test specimen (strain rate of $10^{-3}$ $s^{-1}$) pre-charged with hydrogen. The crack propagates initially from the edge of the specimen towards the center (along the thickness), as shown by the red arrow. Thereafter, the crack propagation occurs from one end of the tensile specimen to the other (indicated by a red dotted arrow). Based on the chevron markings, it can be inferred that the crack propagates from the center (along the thickness) towards the edge of the specimen; it is highlighted by the two smaller red arrows. (b) Fractograph of the crack initiation area and immediate crack propagation radially. (c) Higher magnification micrograph of the crack initiation region; we observe intergranular (black arrow) and quasi-cleavage fracture. (d) and (e) show fracture surface upon significant cracking propagation from one end of the tensile specimen to the other. (d) At the center of the specimen (along thickness), the fracture mode was predominantly brittle. (e) Adjacent to the shear lip, local failure indicates mixed-mode fracture comprising both ductile and brittle fracture features.*

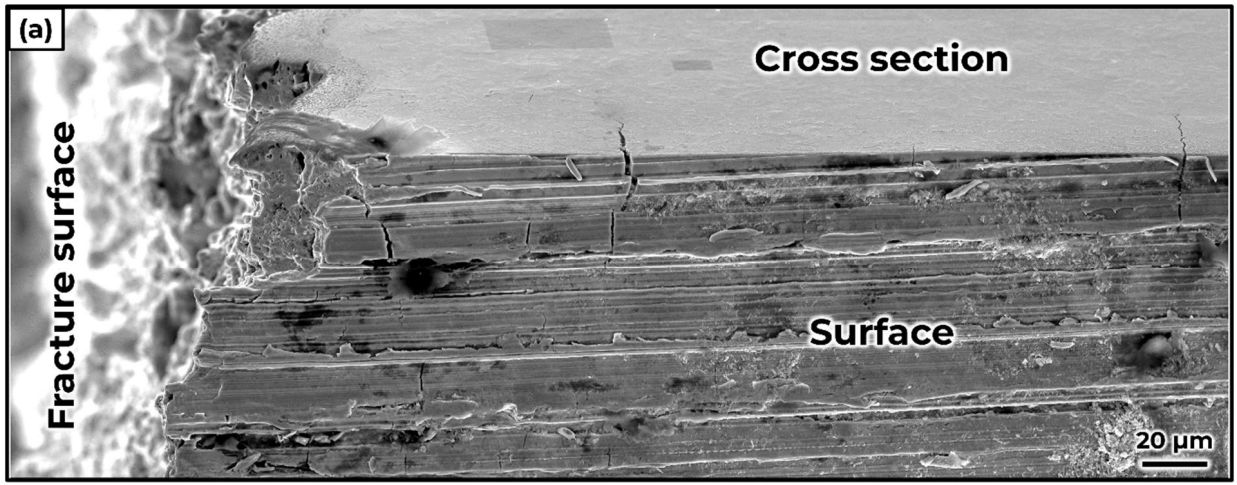

***Figure S2*** *(a) Prescence of multiple edge and surface cracks. (b-c) Discontinuous cracking: formation of fresh crack ahead of the crack tip.*

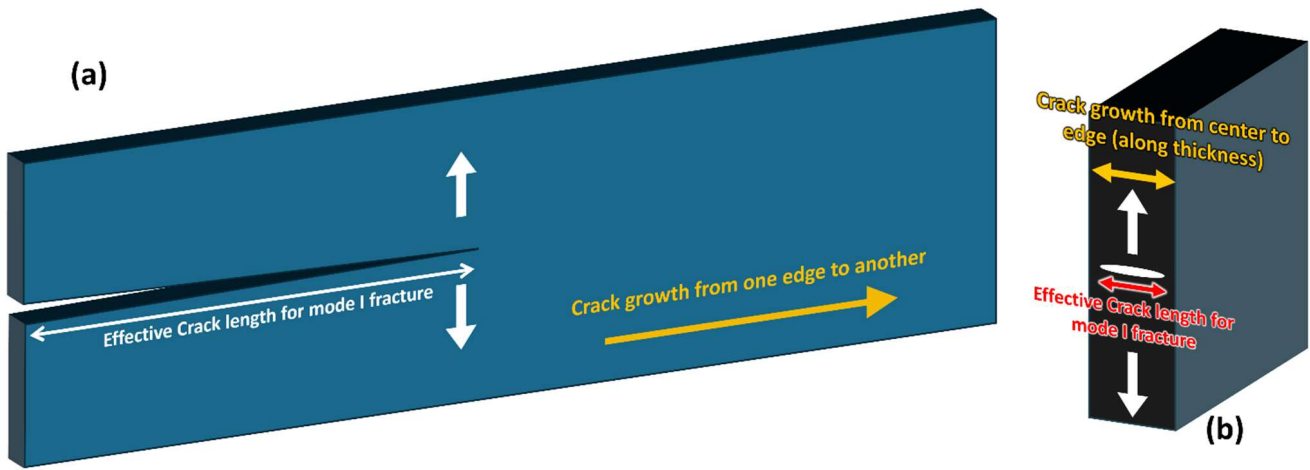

*Figure S3 Schematic of effective crack length for mode I fracture when the crack propagates from (a) one end to another and (b) center to the edge of the specimen along the thickness.*